\documentclass[conference]{IEEEtran}

\ifCLASSINFOpdf
   \usepackage[pdftex]{graphicx}
   \DeclareGraphicsExtensions{.pdf,.jpeg,.png}
\else
   \usepackage[dvips]{graphicx}
   \DeclareGraphicsExtensions{.eps}
\fi
\usepackage[cmex10]{amsmath}
\begin{document}
%
\title{Charge-Focusing Readout of Time Projection Chambers}

\author{\IEEEauthorblockN{S.J. Ross, M.T. Hedges, I. Jaegle, M.D. Rosen, I.S. Seong, T.N. Thorpe, S.E. Vahsen, J. Yamaoka}
\IEEEauthorblockA{University of Hawaii\\
Honolulu, HI 96822\\
}
}


%


\maketitle

\begin{abstract}
Time projection chambers (TPCs) have found a wide range of applications in particle physics, nuclear physics, and homeland security. For TPCs with high-resolution readout, the readout electronics often dominate the price of the final detector. We have developed a novel method which could be used to build large-scale detectors while limiting the necessary readout area. By focusing the drift charge with static electric fields, we would allow a small area of electronics to be sensitive to particle detection for a much larger detector volume. The resulting cost reduction could be important in areas of research which demand large-scale detectors, including dark matter searches and detection of special nuclear material. We present simulations made using the software package Garfield of a focusing structure to be used with a prototype TPC with pixel readout. This design should enable significant focusing while retaining directional sensitivity to incoming particles. We also present first experimental results and compare them with simulation.
\end{abstract}


%
\IEEEpeerreviewmaketitle

\section{IEEE Copyright Notice}
Copyright 2012 IEEE. Published in the conference record of the 2012 IEEE Nuclear Science Symposium and Medical Imaging Conference, October 29-November 3, 2012, Anaheim, CA, USA. Personal use of the material is permitted. However, permission to reprint/republish this material for advertising or promotional purposes or for creating new collective works for resale or redistribution to servers or lists, or to reuse any copyrighted component of this work in other works, must be obtained from the IEEE.

\section{Introduction}
We are working to apply charge focusing to an existing TPC which uses Gas Electron Multipliers (GEMs) to amplify the drift charge and the ATLAS pixel chip to detect it. The idea for charge focusing was originally proposed by Sven Vahsen (University of Hawaii) and John Kadyk (Lawrence Berkeley National Laboratory). Pixel electronics have the advantage of drastically reducing detector noise, which scales with the capacitance of each detector cell and thus also scales with the cell area. Pixels also have excellent timing performance (sampling at 40 MHz), which determines the resolution in the drift direction. The combination of pixel readout and charge focusing allows us to retain the advantages of pixels while instrumenting a large detector. Additionally the increased charge density due to focusing gives higher readout efficiency at constant threshold.

In developing the charge focusing idea, there were two main questions we needed to address. First, could we make this focusing homogeneous? This is desirable as it makes the detector response uniform across the pixel chip. Second, could we limit charge diffusion during focusing? Low levels of diffusion are necessary for retaining directional sensitivity when reconstructing the short tracks expected from WIMP and neutron recoils. Since the focused track will be even shorter, we must keep the diffusion below that of the rest of the detector divided by the focusing factor. 



\section{Initial Simulation}
\begin{figure}
\centering
\includegraphics[width=\linewidth]{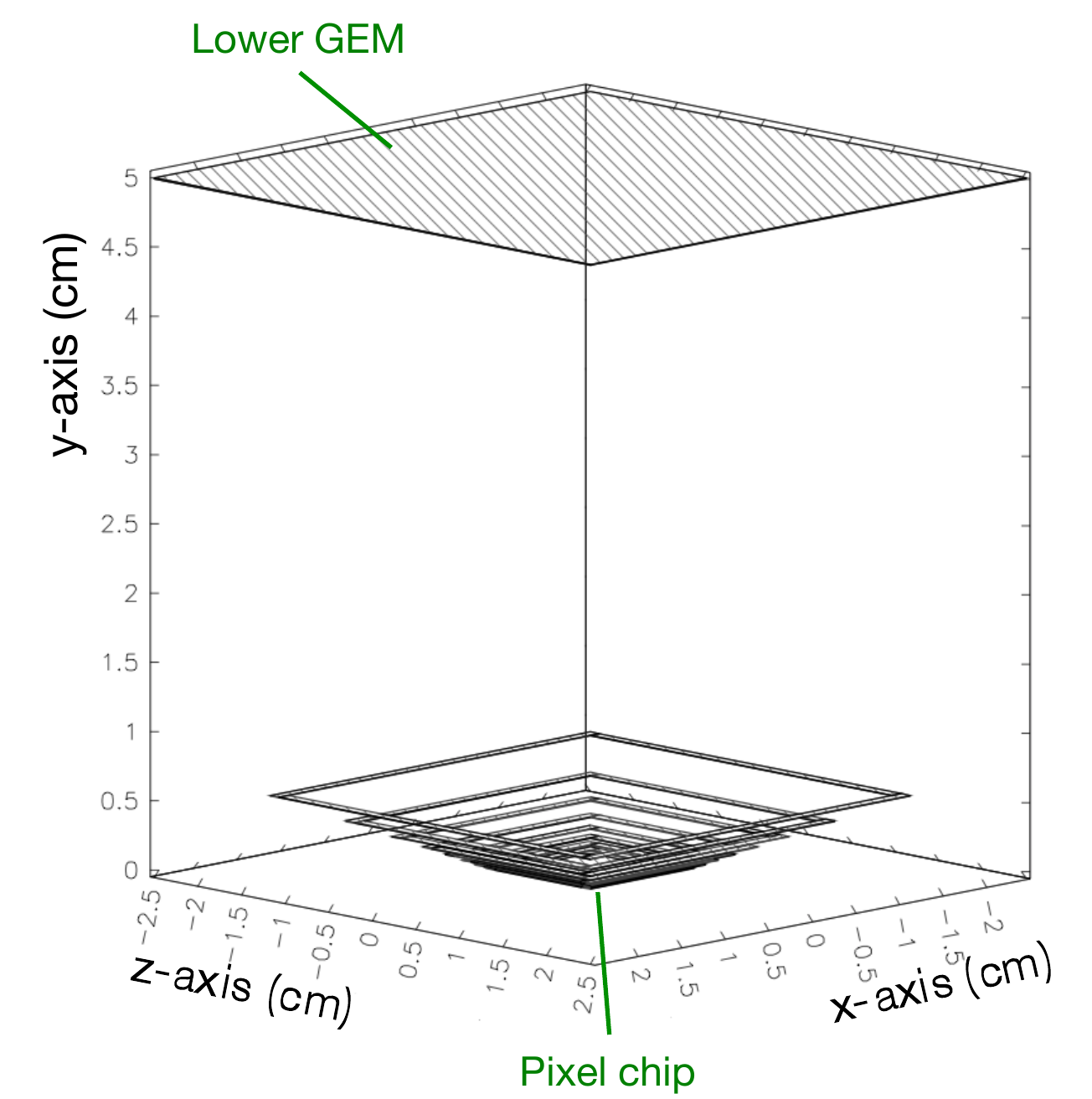}
\caption{Proposed focusing geometry}
\label{waveguidegeometry}
\end{figure}

Simulations were performed using the Garfield drift program developed at CERN. We have assumed 5 cm$^2$ GEMs and a 1 cm$^2$ pixel chip. The proposed focusing geometry is simulated as one GEM and one pixel chip, each represented as a .005 cm thick box held at a set potential, as well as a series of square rings arranged around the pixel chip. These rings act as electrodes, each held at a set voltage.
In order to produce our desired electric field, we have placed the rings in a pattern somewhat reminiscent of a waveguide and used a 5 cm spacing and 1.4 kV voltage differential between GEM and pixel chip (see Figure \ref{waveguidegeometry}). 

To simulate the paths of drifting electrons, we used Garfield's built-in Runge-Kutta-Fehlberg (RKF) solver. The RKF method gives the average expected path for a single electron drifting from a certain point in the detector, and allows a relatively quick assessment of the general drift characteristics of a particular geometry as compared to performing a full Monte Carlo simulation. We simulated electrons drifting from 26 different starting points arranged in a line running along the x-axis from the center of the GEM to one of its edges. Comparing the initial and final
x-positions of the electrons allowed us to calculate an approximate "focusing factor" for each starting point (i.e. an initial track which is contracted to 1/5 its length at readout will have a focusing factor of 5). 
\begin{figure}
\centering
\includegraphics[width=\linewidth]{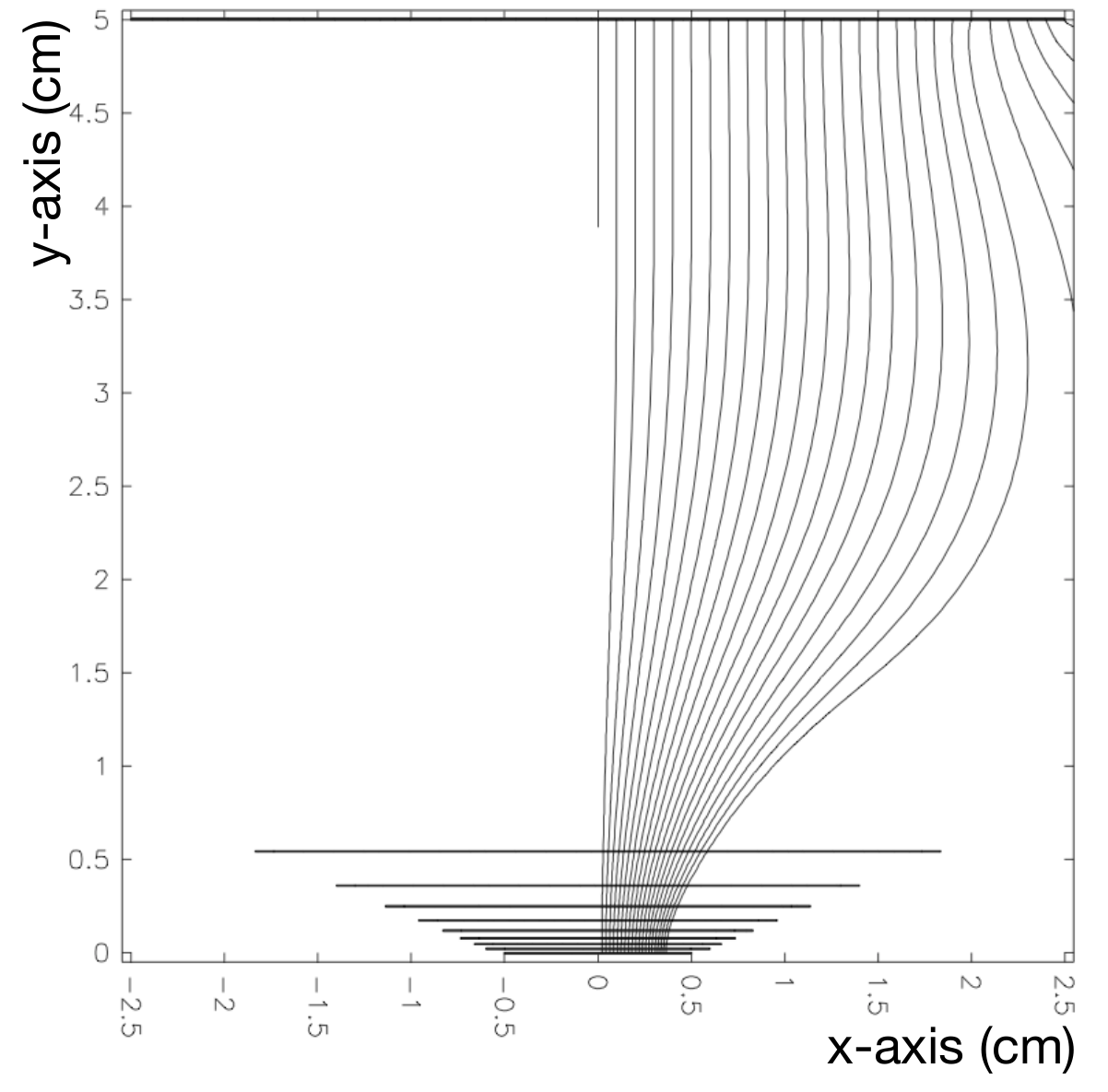}
\caption{Electron drift lines determined with RKF method}
\label{driftlines}
\end{figure}
\begin{figure}
\centering
\includegraphics[width=\linewidth]{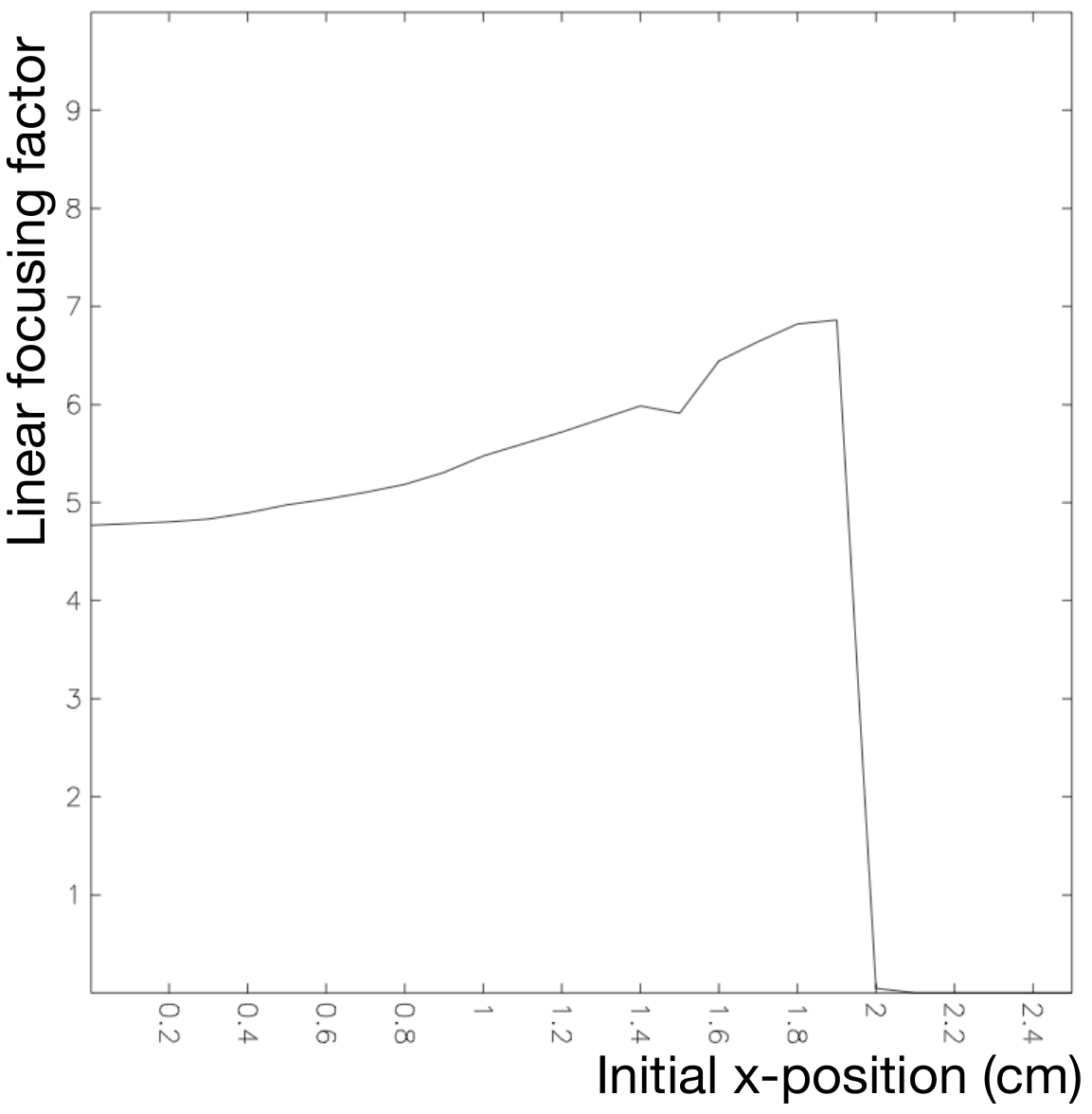}
\caption{Linear focusing factor along x-direction}
\label{focusingfactor}
\end{figure}
\begin{figure}
\centering
\includegraphics[width=\linewidth]{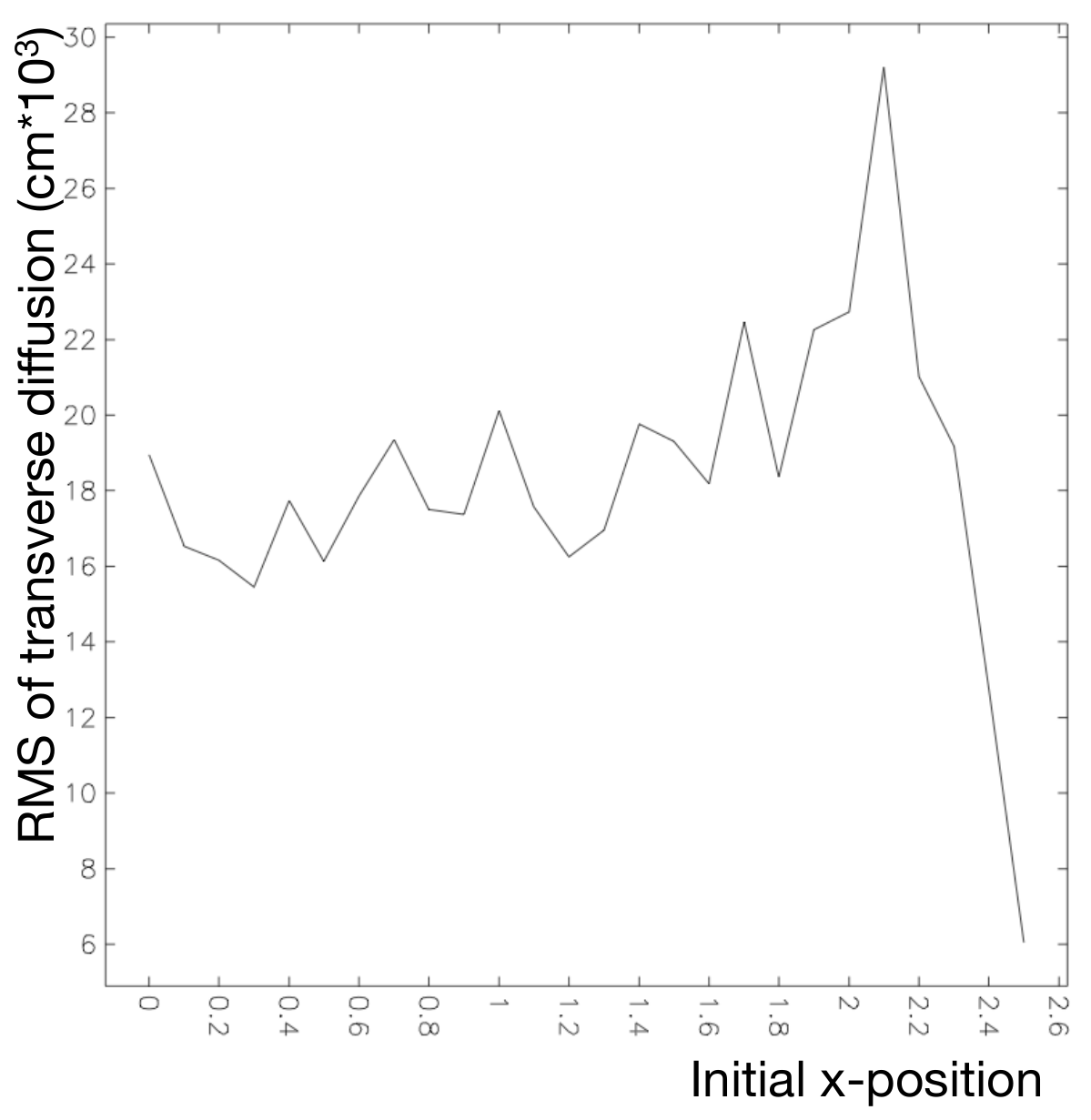}
\caption{RMS values of the transverse diffusion (z-direction)}
\label{RMS}
\end{figure}According to this simulation, our design should give us a fairly consistent focusing factor throughout a large part of the drift region (see Figures \ref{driftlines} and \ref{focusingfactor}). 

We also wanted to ensure that we could limit diffusion of the electrons during focusing. To measure the transverse diffusion, we performed Monte Carlo drift simulations of groups of electrons from the same starting positions as those used in our previous RKF drift simulation. The results of this simulation are shown in Figure \ref{RMS}. We found that, for the majority of the chip, the diffusion was on the order of 200 microns.

\section{Experimental Verification}
Our initial simulations were promising, but we wanted to make sure they were accurate. One concern was the fact that Garfield does not account for inter-electron Coulomb repulsion.

\begin{figure}[b]
\centering
\includegraphics[width=\linewidth]{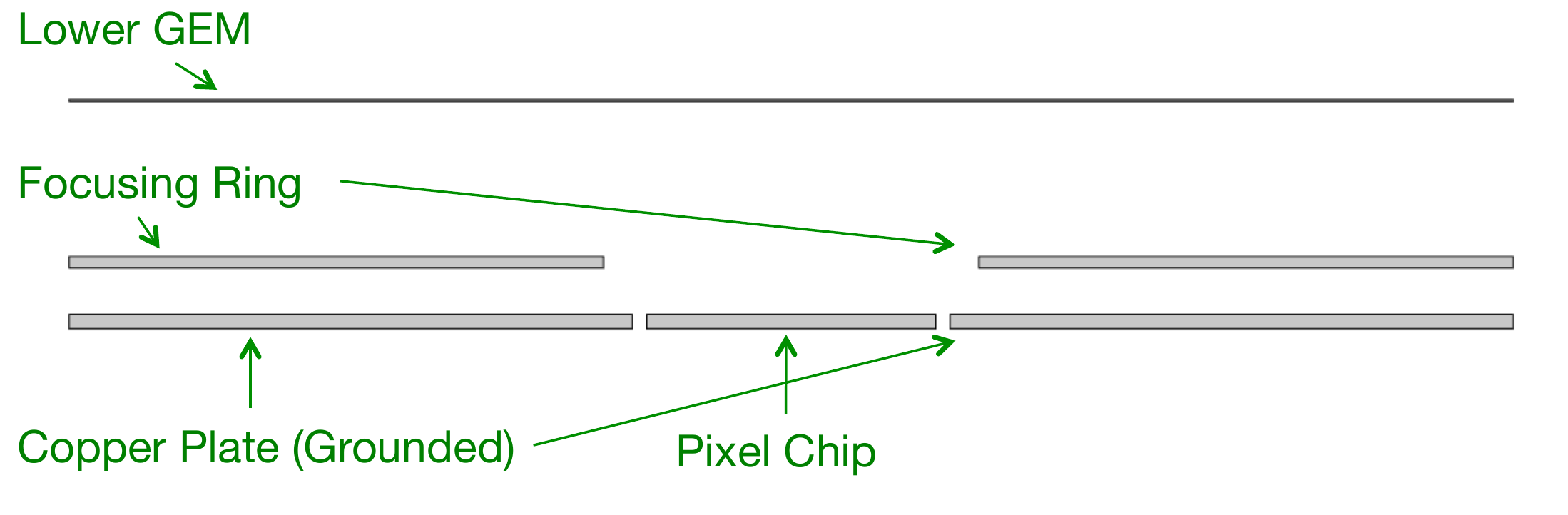}
\caption{2D cross-sectional view of test setup}
\label{labtestgeometry}
\end{figure}
We decided to create a simplified test setup to determine the validity of our simulations. This design consists of a single, square ring placed between the GEMs and pixel chip, as shown in Figure \ref{labtestgeometry}. By adjusting the voltage on the ring, we can switch between focusing and non-focusing modes. 
\begin{figure}
\centering
\includegraphics[width=\linewidth]{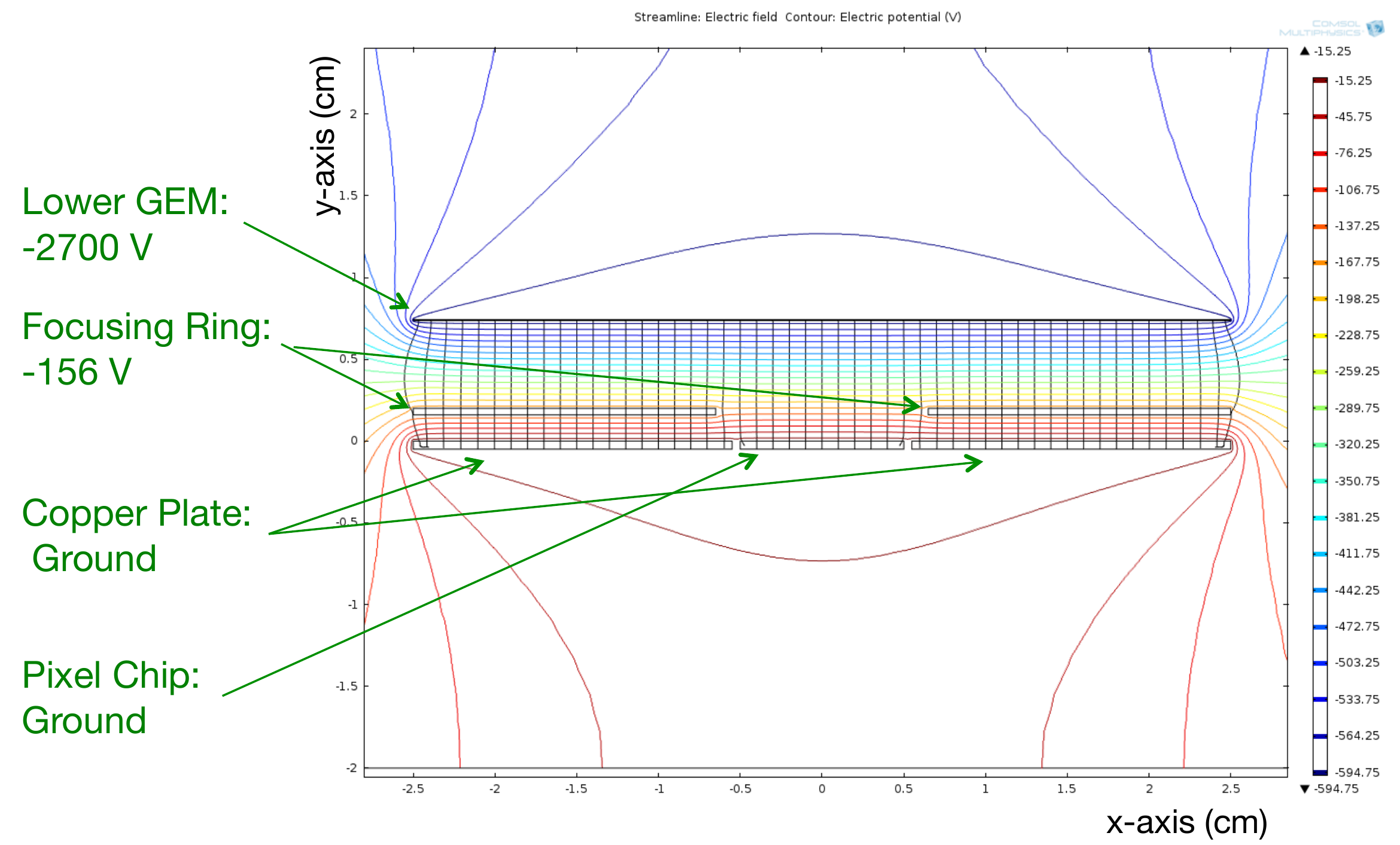}
\caption{COMSOL simulation of non-focusing test setup}
\label{comsolnonfocusing}
\end{figure}
\begin{figure}
\centering
\includegraphics[width=\linewidth]{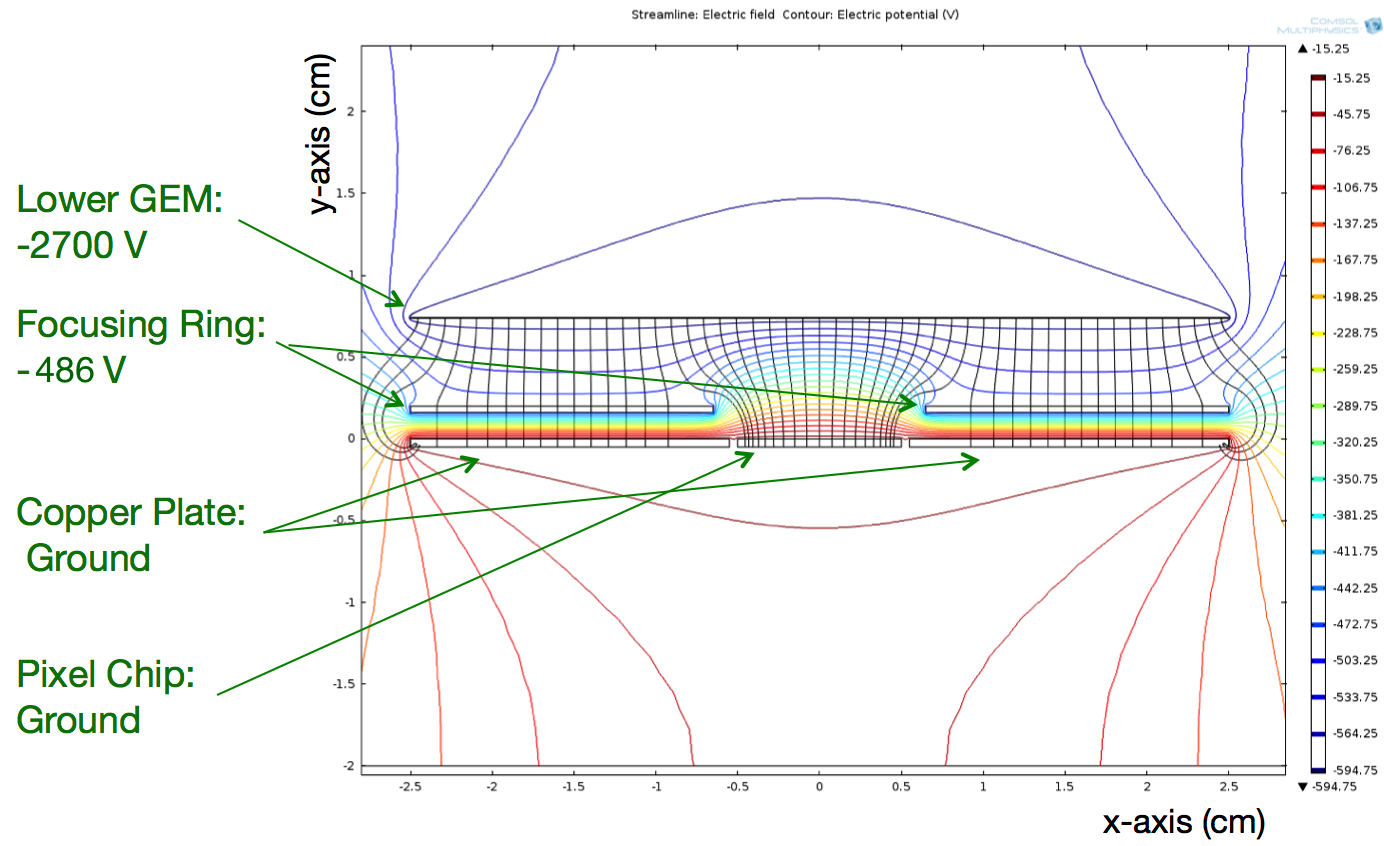}
\caption{COMSOL simulation of focusing test setup}
\label{comsolfocusing}
\end{figure}These two modes, simulated with COMSOL Multiphysics, are pictured in Figures \ref{comsolnonfocusing} and \ref{comsolfocusing}.

\subsection{Drift Simulation}
To simulate electron drift with this simplified geometry, we again built a 3D simulation of the geometry using Garfield. As before, we used the Runge-Kutta-Fehlberg technique to determine the drift lines for electrons with starting positions along the x-axis. For this simulation, we used 250 evenly-spaced starting points and drifted in a 70\% Argon, 30\% CO$_2$ gas mixture at 1 atm. 
\begin{figure}
\centering
\includegraphics[width=\linewidth]{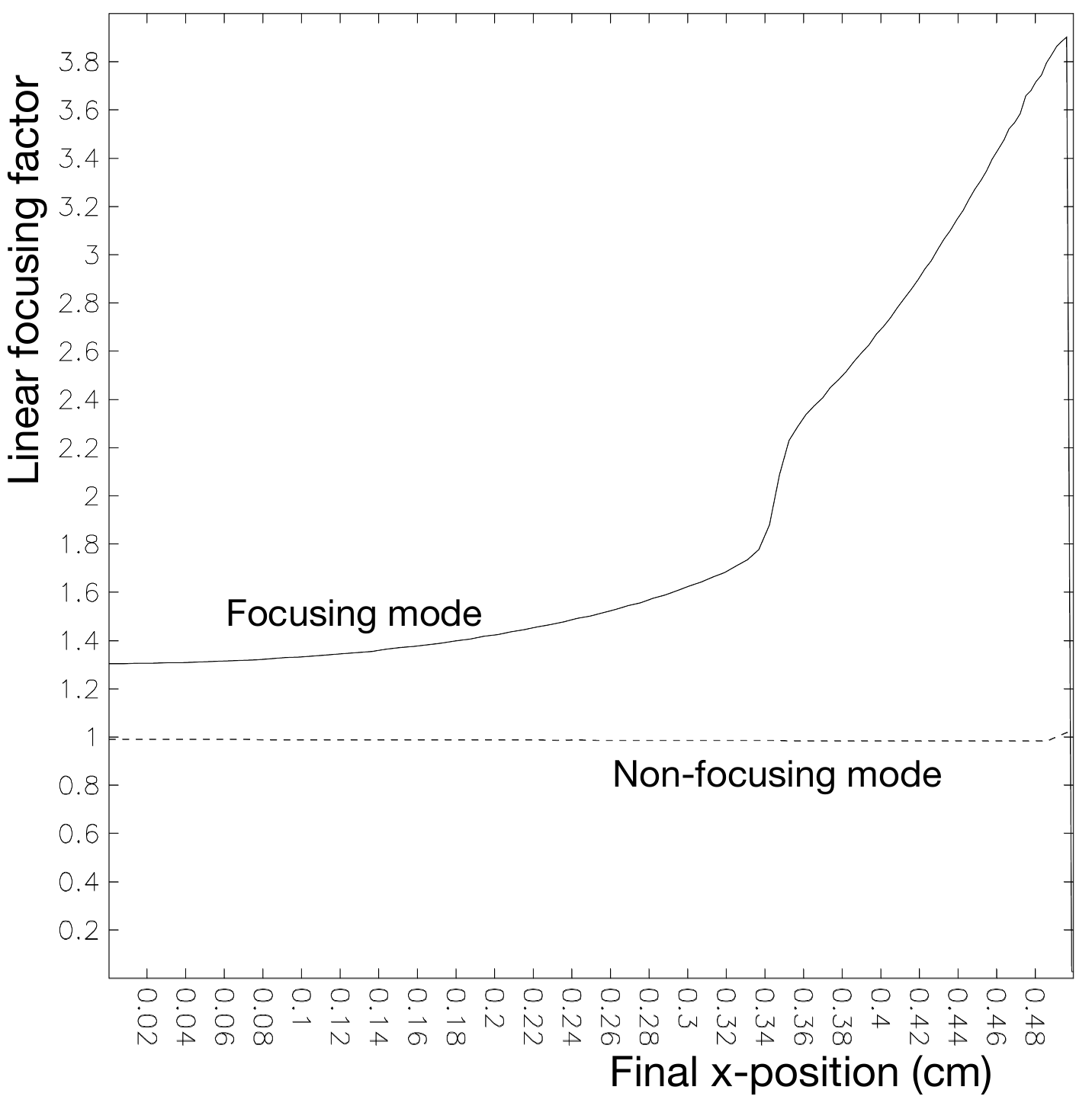}
\caption{Linear focusing factor along x-direction}
\label{labtestfocusingfactor}
\end{figure}The focusing factor values across the extent of the pixel chip are shown in Figure \ref{labtestfocusingfactor} This focusing is clearly non-homogeneous, but it should give us a measurable difference between the focusing and non-focusing modes.

In order to predict the behavior of this experimental setup, we simulated the difference in the electron hit rate that should exist between the focusing and non-focusing modes. 
\begin{figure}
\centering
\includegraphics[width=\linewidth]{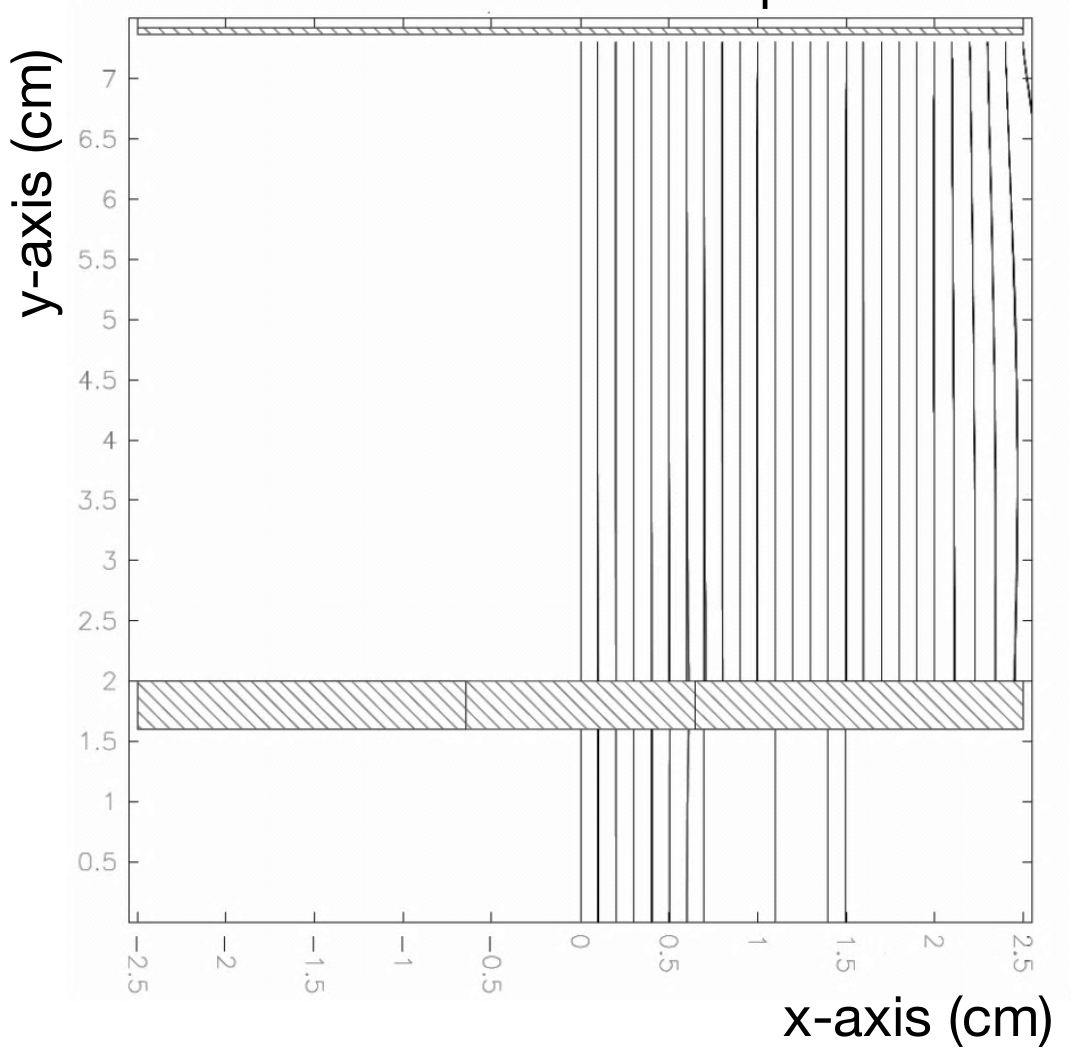}
\caption{RKF simulation of drift lines originating across one quadrant of GEM. Here focusing is turned off.}
\label{nonfocusingrateRKF}
\end{figure}
\begin{figure}
\centering
\includegraphics[width=\linewidth]{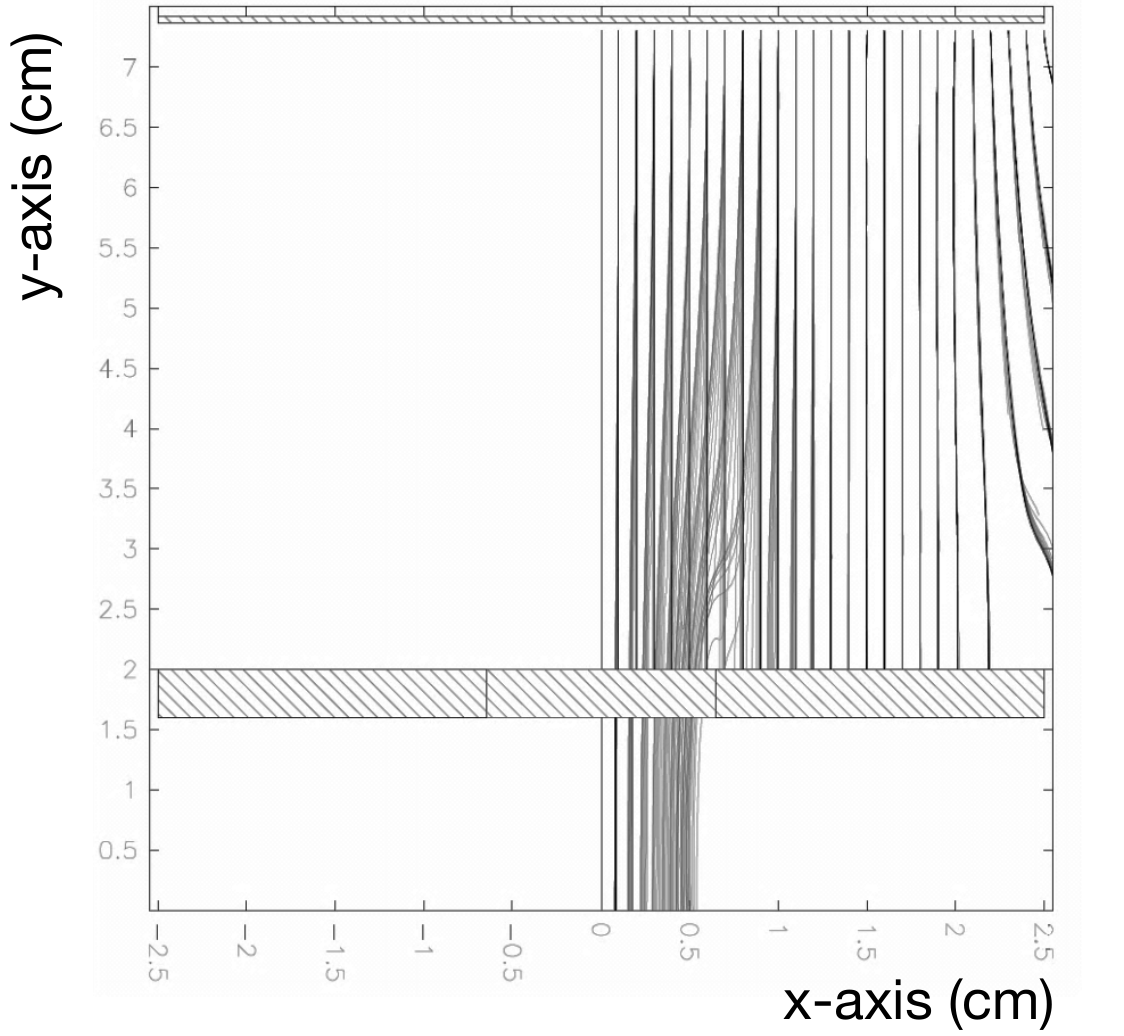}
\caption{RKF simulation of drift lines originating across one quadrant of GEM. Here focusing is turned on.}
\label{focusingrateRKF}
\end{figure}To do so, we simulated RKF drift lines originating at evenly-spaced intervals across one quadrant of the GEM, as shown in Figures \ref{nonfocusingrateRKF} and \ref{focusingrateRKF}. We then counted the number of drift lines ending on the pixel chip. This should be equivalent to the hit rate assuming the ionization is produced by radiation that is homogeneously distributed throughout the detector volume. We found that 83 drift lines terminated on the chip with focusing turned on, while only 37 did so with the focusing off. This is equivalent to a 2.24$\times$ rate increase from non-focusing to focusing.

\subsection{Lab Installation}
\begin{figure}
\centering
\includegraphics[width=\linewidth]{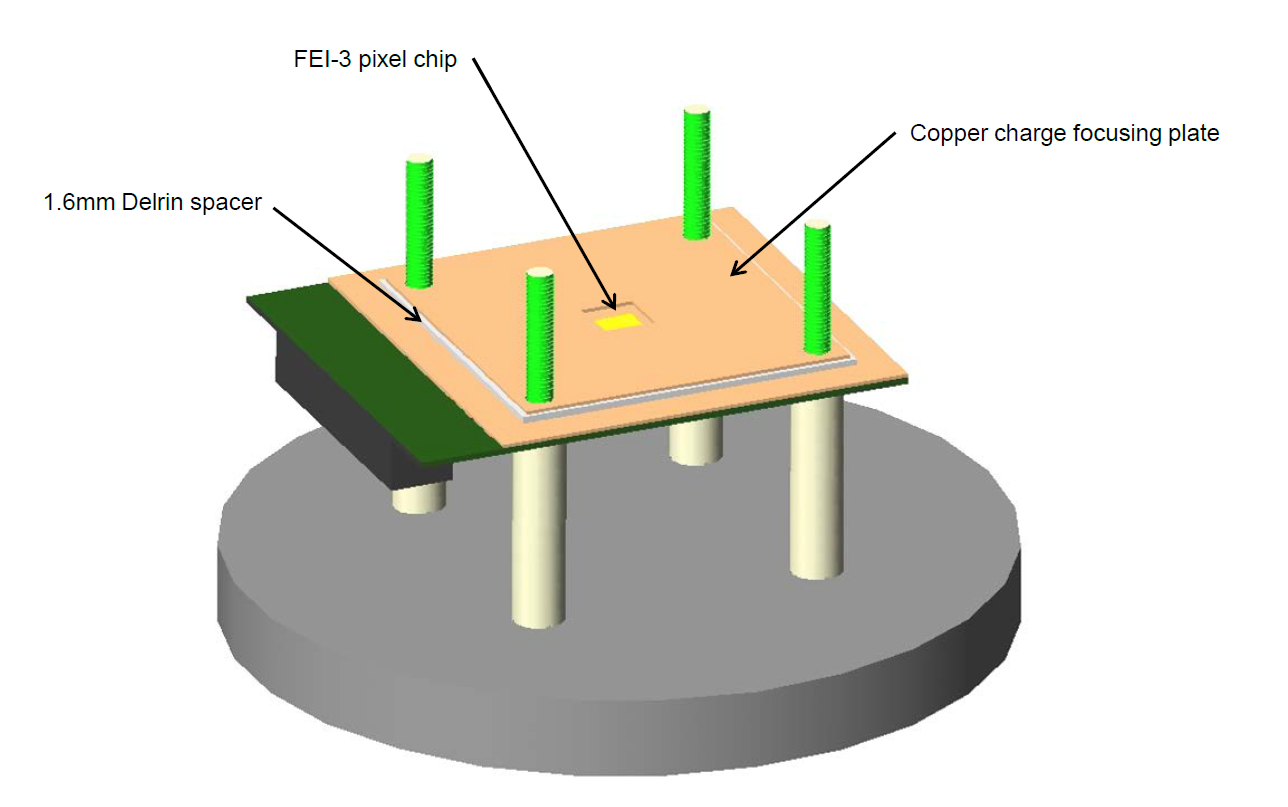}
\caption{CAD drawing of focusing ring and pixel installation}
\label{focusingsetup}
\end{figure}To implement this design in our detector, we cut the focusing ring out of copper sheet and installed it in our TPC as depicted in Figure \ref{focusingsetup}. 
\begin{figure}
\centering
\includegraphics[width=\linewidth]{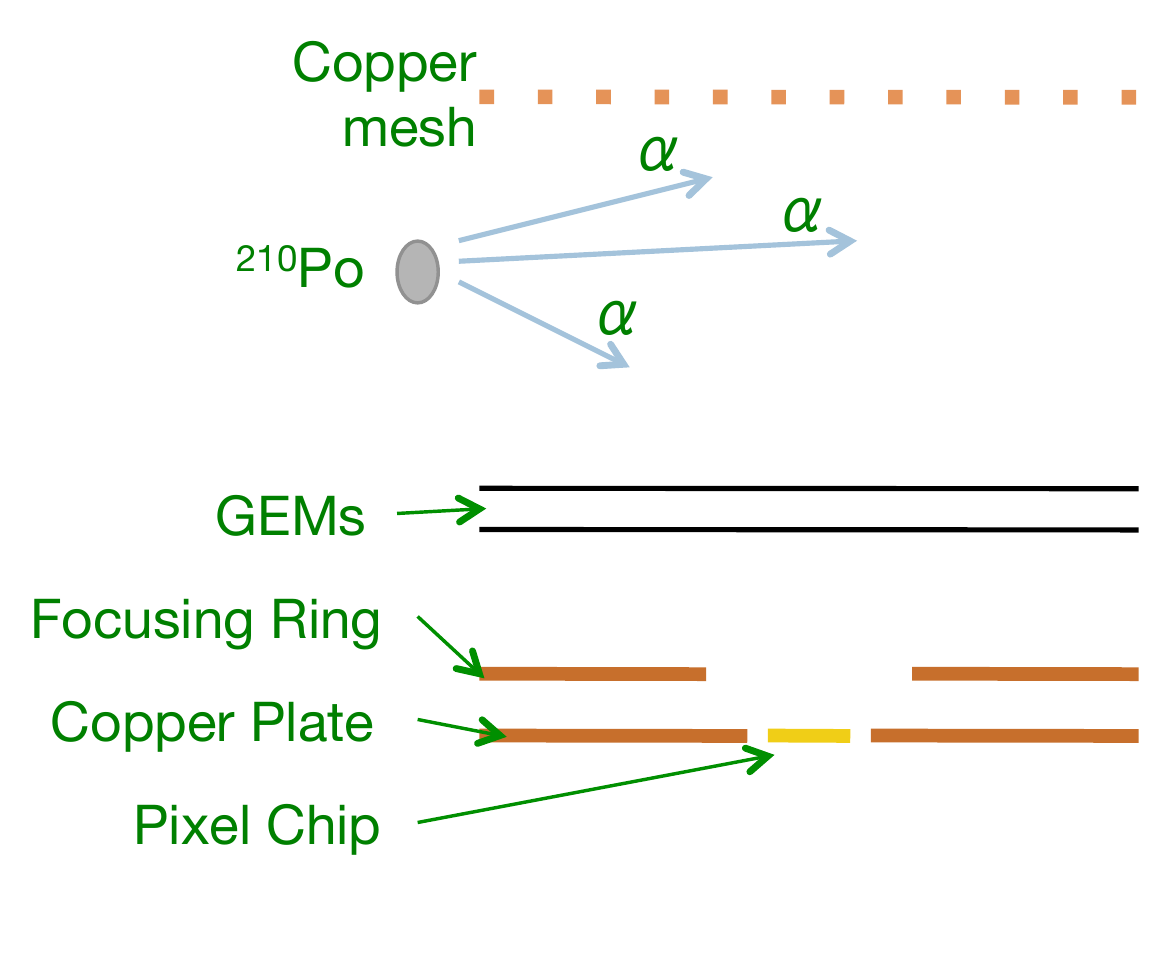}
\caption{Simplified drawing of lab test setup with alpha particle source}
\label{labtestsetup}
\end{figure}We used a Polonium-210 alpha particle source to produce ionization tracks in our detector. This setup is shown in Figure \ref{labtestsetup}.

\subsection{Hit Rate Analysis}
To determine the actual hit rate for our experimental setup, we replaced the pixel chip with a solid copper pad and connected it to a pulse height analyzer. We then measured the number of hits in a given time period with the focusing turned on and off. By fitting the resulting alpha energy peak with a Gaussian, we determined the total hit rate due to alpha particles.
\begin{figure}
\centering
\includegraphics[width=\linewidth]{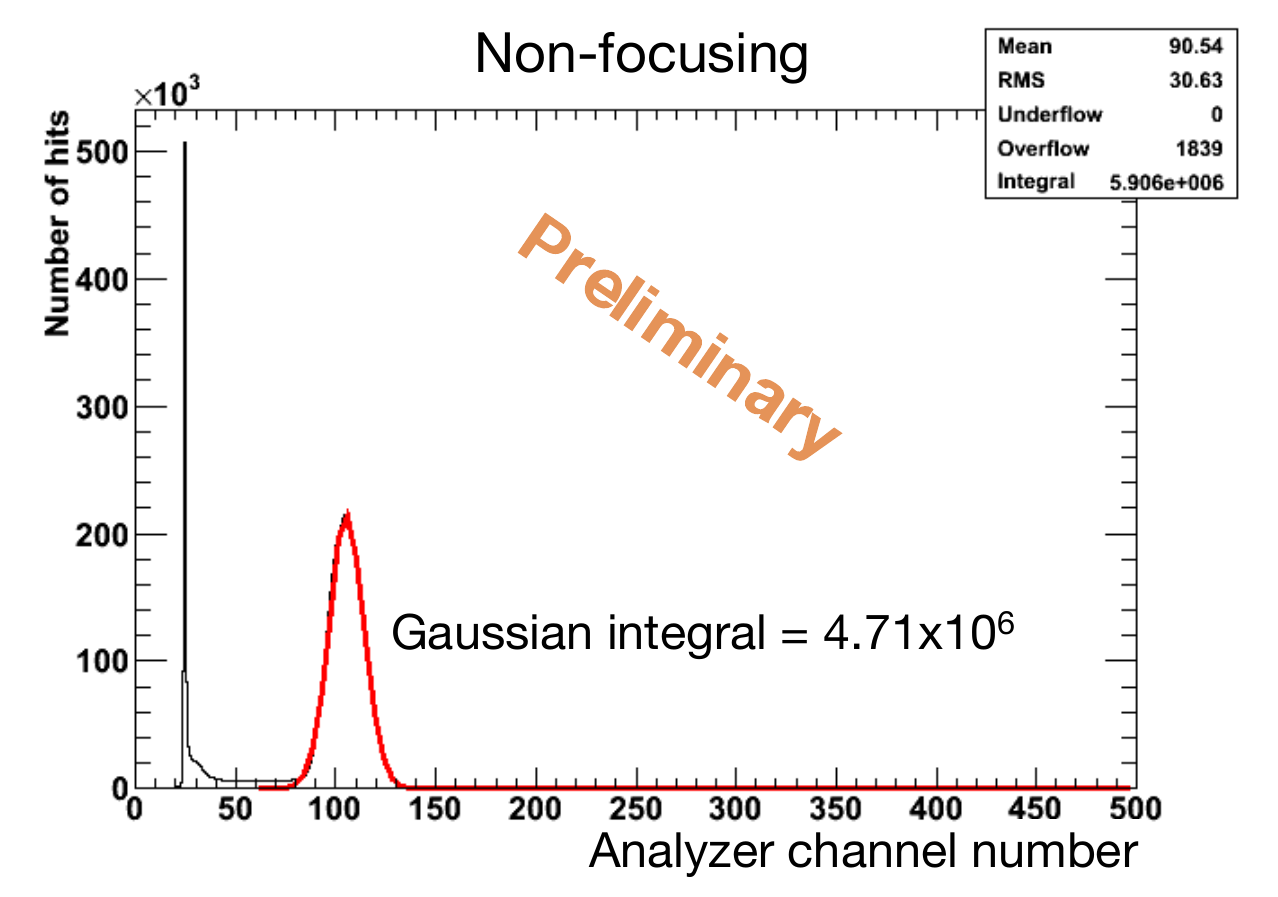}
\caption{Electron hit count due to alpha particles in detector with focusing off. Hits were recorded over 299 seconds.}
\label{nonfocusingrate}
\end{figure}
\begin{figure}
\centering
\includegraphics[width=\linewidth]{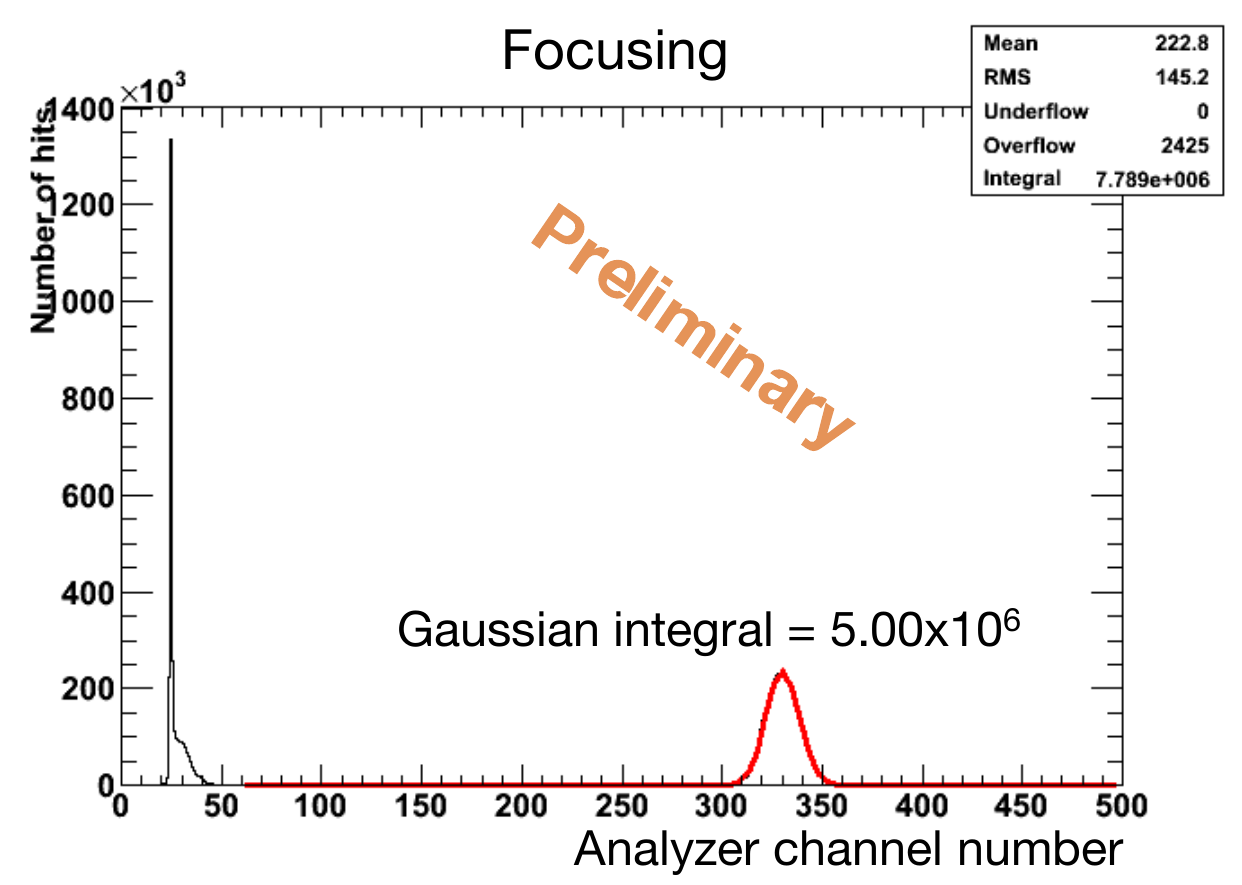}
\caption{Electron hit count due to alpha particles in detector with focusing on. Hits were recorded over 299 seconds.}
\label{focusingrate}
\end{figure}
These results are shown in Figures \ref{nonfocusingrate} and \ref{focusingrate}.

In order to find accurate values for these measured hit rates, we needed to account for the detector dead time, which differs between the focusing and non-focusing runs and affects the amount of active time for which the detector is actually recording data. To find the actual rate, we used the  equation
\begin{equation}R=\frac{R_m}{1-R_m\tau}\cite{deadtime}\end{equation}
where R is the actual rate, R$_m$ is the measured rate, and $\tau$ is the total dead time.

Applying the dead time correction to our measured data, we found a hit rate of 6.04$\times 10^4$hits/s with focusing turned off, and $1.52\times 10^5$hits/s with focusing on, giving a 2.51$\times$ rate increase from non-focusing to focusing. This value is not in perfect agreement with the 2.24$\times$ increase our simulation predicted, but it does support the fact that our design is focusing the ionization tracks of the alpha particles.

\subsection{Preliminary Pixel Data}
After performing the rate analysis with the pulse height analyzer, we installed the pixel chip in the detector and again recorded data for both focusing and non-focusing modes. More work still needs to be done on this data to correctly separate alpha tracks from noise events, but we do have some encouraging preliminary analysis.
\begin{figure}
\centering
\includegraphics[width=\linewidth]{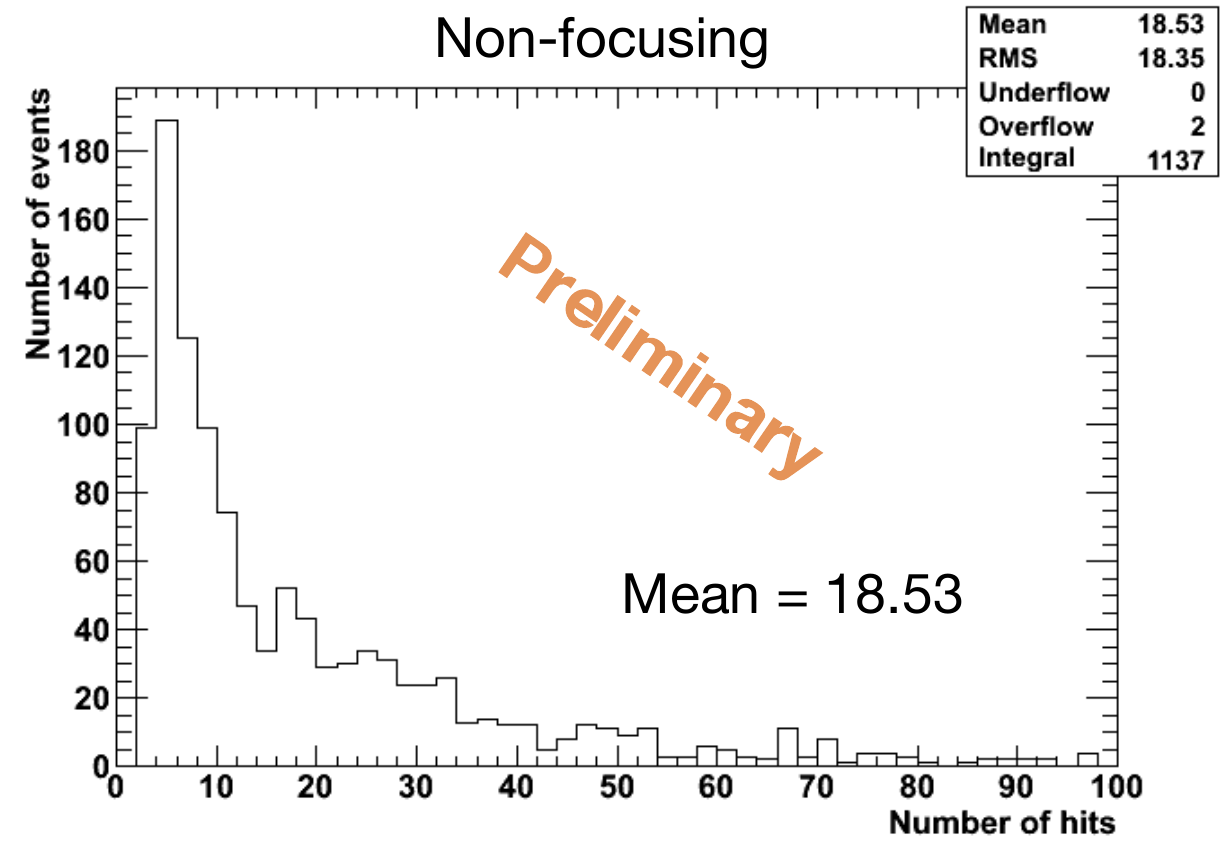}
\caption{Number of pixel hits recorded per event read out by pixel chip with detector in non-focusing mode}
\label{nonfocusingpixel}
\end{figure}
\begin{figure}
\centering
\includegraphics[width=\linewidth]{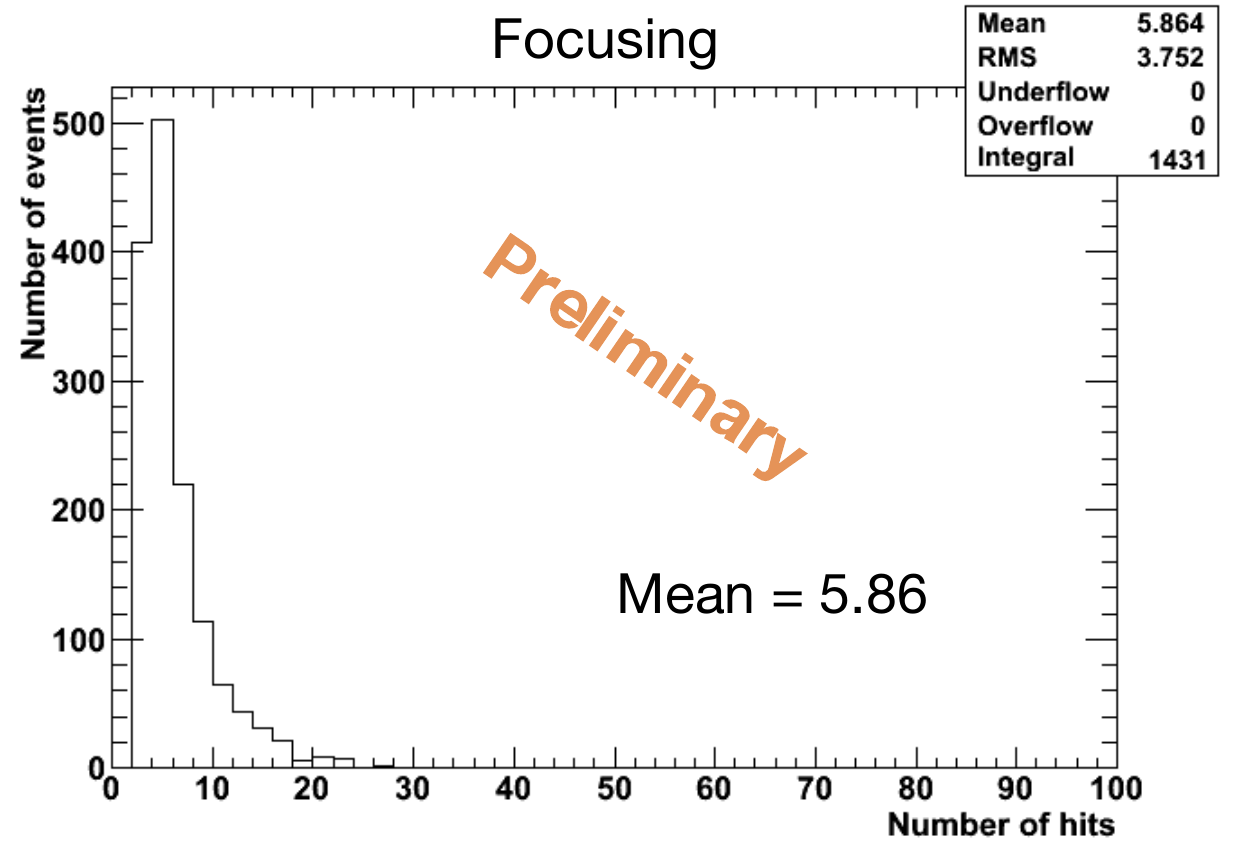}
\caption{Number of pixel hits recorded per event read out by pixel chip with detector in focusing mode}
\label{focusingpixel}
\end{figure}Figures \ref{nonfocusingpixel} and \ref{focusingpixel} show the number of pixel hits recorded for each event that was read out by the pixel chip for the non-focusing and focusing modes, respectively. This is essentially equivalent to a measurement of each event's size. As can be seen from these figures, there is an approximately 3$\times$ event size decrease from non-focusing to focusing. This suggests that with the focusing turned on the electrons from each alpha track are focused onto fewer pixels.

\section{Conclusion}
We have created simulations using Garfield which demonstrate that it should be possible to implement charge focusing while minimizing diffusion and keeping focusing relatively homogeneous across the readout area.

We have also created a simplified experimental setup in order to verify the accuracy of our simulations. Both electron hit rate and preliminary event size analysis show evidence of focusing in our lab setup; we have rough but imperfect agreement with the predictions of our simulation. We will continue to analyze the alpha track data recorded by the pixel chip.

\begin{figure}
\centering
\includegraphics[width=0.8\linewidth]{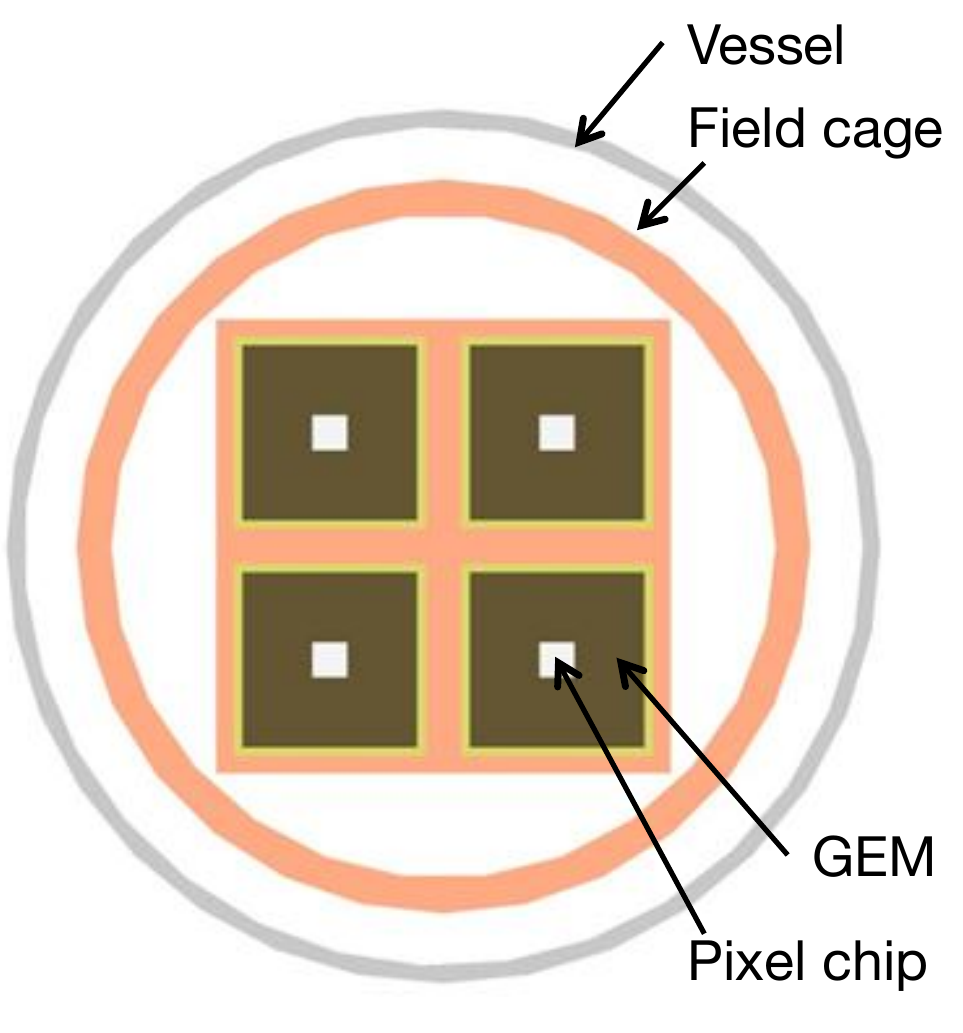}
\caption{Proposed layout for next-generation TPC}
\label{nextgen}
\end{figure}
We plan to further develop charge focusing with the next-generation D$^3$ detector prototype. This detector (as shown in Figure \ref{nextgen}) will incorporate four readout chips. We believe the adjacent cell placement should improve edge effects seen in our simulations.


\section*{Acknowledgment}
We acknowledge fruitful discussions with John Kadyk and Maurice
Garcia-Sciveres of Lawrence Berkeley National Laboratory, and thank
them for providing the ATLAS FE-I3 pixel chip and associated readout
electronics used in this work. We acknowledge support from the U.S.
Department of Homeland Security under Award Number
2011-DN-077-ARI050-03 and the U.S. Department of Energy under Award
Number DE-SC0007852. The views and conclusions contained in this
document are those of the authors and should not be interpreted as
necessarily representing the official policies, either expressed or
implied, of the United States Government or any agency thereof.




\begin{thebibliography}{1}

\bibitem{deadtime}
M. L. Larsen and A. B. Kostinski, "Simple dead-time corrections for discrete series of non-Poisson data," \emph{Meas. Sci. Technol.}, vol. 20, no. 9,  Sept. 2009
\end{thebibliography}
%

\end{document}